\documentclass[conference]{IEEEtran}
\IEEEoverridecommandlockouts
\usepackage{cite}
\usepackage[numbers,sort&compress]{natbib}
\usepackage{amsmath,amssymb,amsfonts}
\usepackage{algorithmic}
\usepackage{graphicx}
\usepackage{textcomp}
\usepackage{xcolor}
\usepackage{url}
\usepackage{bm}
\usepackage{pifont} 

\newcommand{\cmark}{\textcolor{green}{\ding{51}}}
\newcommand{\xmark}{\textcolor{red}{\ding{55}}} 

\usepackage{amsmath}
\usepackage{bm}
\usepackage{multirow,tabularx}
\usepackage{hhline}
\usepackage{mathtools}
\usepackage{subfigure}
\usepackage{multirow,tabularx}
\usepackage{makecell}
\newtheorem{problem}{Problem}

\usepackage{setspace}
\usepackage{xcolor}
\usepackage[normalem]{ulem} 
\usepackage{adjustbox}
\DeclareMathOperator*{\argmax}{arg\,max}
\DeclareMathOperator*{\argmin}{arg\,min}
\usepackage{booktabs}

\def\BibTeX{{\rm B\kern-.05em{\sc i\kern-.025em b}\kern-.08em
    T\kern-.1667em\lower.7ex\hbox{E}\kern-.125emX}}

\begin{document}

\title{Bi-NAS: Towards Effective and Personalized Explanation for Recommender Systems via Bi-Level Neural Architecture Search\\}


\author{%
\setlength{\tabcolsep}{12pt}
\renewcommand{\arraystretch}{1.0}
\begin{tabular}{@{}cccc@{}}

\begin{tabular}[t]{@{}c@{}}
Longfeng Wu \\
\textit{Virginia Tech}\\
Blacksburg, VA, USA \\
longfengwu@vt.edu
\end{tabular} &
\begin{tabular}[t]{@{}c@{}}
Yao Zhou\\
\textit{Google}\\
Mountain View, CA, USA\\
yaozhoucosmos@google.com
\end{tabular} &
\begin{tabular}[t]{@{}c@{}}
Tong Zeng\\
\textit{Virginia Tech}\\
Blacksburg, VA, USA\\
tongzeng@vt.edu
\end{tabular} &
\begin{tabular}[t]{@{}c@{}}
Zhimin Peng\\
\textit{Amazon}\\
Sunnyvale, CA, USA\\
zmpeng@amazon.com
\end{tabular}
\\[5.0em] 

\begin{tabular}[t]{@{}c@{}}
Bhanu Pratap Singh Rawat\\
\textit{Amazon}\\
Sunnyvale, CA, USA\\
rawabhan@amazon.com
\end{tabular} &
\begin{tabular}[t]{@{}c@{}}
Lecheng Zheng\\
\textit{Virginia Tech}\\
Blacksburg, VA, USA\\
lecheng@vt.edu
\end{tabular} &
\begin{tabular}[t]{@{}c@{}}
Giovanni Seni\\
\textit{Amazon}\\
Sunnyvale, CA, USA\\
gseni@amazon.com
\end{tabular} &
\begin{tabular}[t]{@{}c@{}}
Dawei Zhou\\
\textit{Virginia Tech}\\
Blacksburg, VA, USA\\
zhoud@vt.edu
\end{tabular}

\end{tabular}%
}


\maketitle

\begin{abstract}
Recommender systems are vital in helping users navigate vast amounts of information, offering personalized suggestions and effective explanations for these recommendations. While previous efforts have attempted to provide such explanations, evaluating their effectiveness across various scenarios remains a challenge. Enhancing these explanations is essential for improving user engagement, trust, and decision-making. 
To facilitate effective explanations within the recommender system, we propose a Bi-level Neural Architecture Search (Bi-NAS) framework to optimize explanations. This approach simultaneously refines cross-attention mechanisms and feature interaction functions by exploring both intra-layer and inter-layer design spaces. 
Furthermore, we integrate Large Language Models (LLMs) to enhance explanation generation, leveraging zero-shot prompting to produce more effective and personalized justifications. By aligning user feature preferences with item quality scores, our approach ensures that explanations reflect both user intent and item attributes, improving transparency and reasoning depth.
Extensive evaluations on four real-world datasets demonstrate that Bi-NAS not only boosts recommendation accuracy but also significantly improves the effectiveness of explanations for recommender systems, providing users with clear and reliable insights into the suggestions they receive. Meanwhile, we publish our data and code at \url{https://github.com/wulongfeng/Bi-NAS.git}.
\end{abstract}

\begin{IEEEkeywords}
Explainable Recommendation, Neural Architecture Search
\end{IEEEkeywords}

\section{Introduction}




In the digital age, with the explosion of information and data, recommender systems have emerged as an essential tool for dealing with information overload~\cite{aggarwal2016recommender}. These systems are designed to provide users with personalized and relevant content, which can help cater to their preferences and needs, stream their decision-making process, and improve service efficiency~\cite{wu2022towards}. 
However, users may not be satisfied with the results they receive without clear and meaningful explanations, as shown in Figure 1. Providing effective explanations for the recommendation results can significantly enhance the users' experience, foster their engagement and trust, and boost their loyalty and stickiness to the service and products. Meanwhile, an effective explanation could also be useful for the users to improve their decision-making accuracy and efficiency~\cite{balog2020measuring} or increase users' satisfaction with the explanations~\cite{xian2021ex3}.

\begin{figure}[thb]
\centering
\includegraphics[width=.48\textwidth]{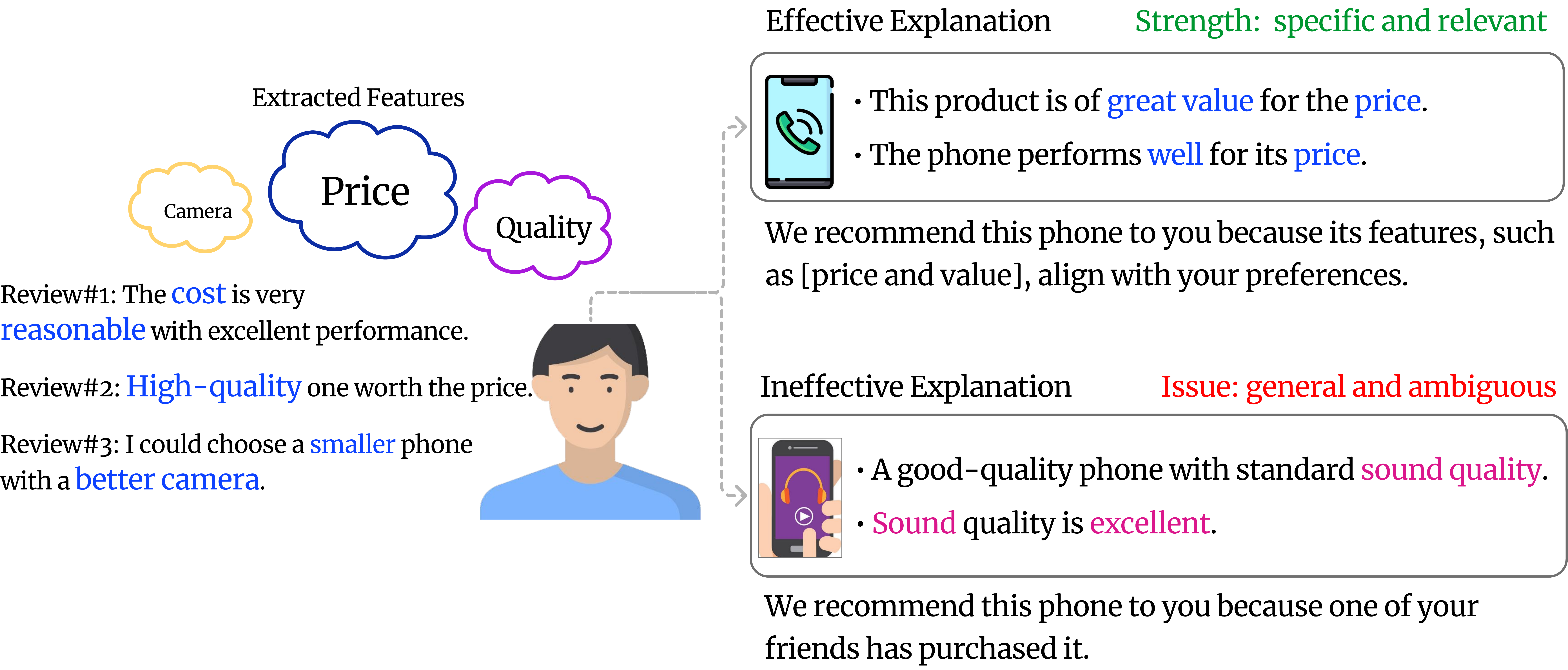}
\caption{Illustration of recommender system with effective explanation. The reviews on the left are written by one user with their key aspects and opinions highlighted in blue. Similarly, each item will also receive multiple reviews from various users, with its key aspects and opinions marked in pink or blue. The word clouds are generated based on the user's perspectives, with each word's size in the word cloud reflecting its frequency, as the user mentioned.}
\label{fig: motivation example}
\end{figure}

Early explainable methods, such as collaborative filtering (CF), generated simple justifications like ``users who liked X also liked Y"~\cite{schafer2007collaborative, koren2009matrix}. While intuitive, these explanations often lack the granularity and persuasive power to be truly compelling. A significant step towards building more interpretable models came from leveraging the rich, explicit information contained in user reviews. Reviews offer direct evidence of user preferences and item characteristics at a fine-grained level. 
As illustrated in Figure~\ref{fig: motivation example}, users' preferences can be discerned through the reviews they compose. By examining the customer's previous reviews, we can deduce that this particular customer places a higher priority on ``price" and ``quality", followed by the camera's attributes. Similarly, we can infer the quality of items by examining the reviews they have received. For instance, the foremost considerations for the first phone revolve around ``price" and ``quality". Consequently, we can recommend items to users by aligning their preferences with the features of the items, thereby offering more insightful and relevant explanations. In contrast, the explanation that one of your friends has purchased is not sufficiently convincing, thus failing to provide effective explanations.
Building on this insight, researchers have developed sophisticated models that incorporate textual data. Attention-based methods~\cite{dong2020asymmetrical, barkan2020explainable, wang2020attention} were introduced to dynamically weigh the importance of different words, sentences, or reviews, allowing the model to focus on the most salient information. Similarly, path-based models~\cite{chen2021temporal, geng2022path,zheng2022explainable}, often leveraging knowledge graphs, aimed to find explicit reasoning paths between users and items. More recently, the advent of Large Language Models (LLMs) has introduced powerful capabilities for natural language understanding and generation, opening new frontiers for creating fluent, human-like explanations.

Despite this progress, generating effective and personalized explanations in recommender systems remains a challenging task due to several critical limitations in existing methods.
First, (\textbf{C1. Heavy reliance on human effort and expertise}) Designing interpretable components, such as explainable paths or attention mechanisms, often requires significant manual effort and domain expertise. This process can be labor-intensive and prone to human bias, which may result in sub-optimal or inconsistent explanations.
Second, (\textbf{C2. Poor generalization across dynamic datasets}) Recommendation tasks often span diverse datasets with varying characteristics. A single, fixed explainability architecture may fail to generalize effectively across these datasets, limiting its adaptability to different recommendation contexts.
Third, (\textbf{C3. Risk of hallucination in LLM-based methods}) While LLMs have shown impressive capabilities in natural language tasks, directly applying them to explanation generation can introduce hallucinations, factual errors, and generic outputs. Without careful adaptation to user-specific contexts, LLMs often fail to capture nuanced preferences, resulting in unreliable and impersonal explanations that undermine user trust.

To address the aforementioned challenges, we propose a Bi-level Neural Architecture Search (Bi-NAS) to jointly search for a recommender system design with optimal, effective, and explainable architecture. 
In particular, to address \textbf{C1} and \textbf{C2}, we introduce NAS, which aims to automatically find a neural architecture that achieves the best possible performance with limited computing resources while requiring minimal human intervention~\cite{ren2021comprehensive}. NAS strategically explores possible neural architectures within a predefined search space and adapts to different datasets, thus generalizing to diverse datasets by producing tailored architectures for varying data characteristics.
Additionally, to address \textbf{C3}, we employ LLMs to enhance explanation generation using aligned features between users and items discovered by NAS, while incorporating user history to enable more personalized and contextually relevant explanations.
Our approach learns user preference weights and item quality weights by extracting meaningful features from user reviews. By aligning users' feature preferences with item quality scores on corresponding attributes, we leverage LLMs to generate effective and personalized explanations.  
Specifically, our method first learns the embedding of users, items, and their associated reviews. Next, we employ Bi-NAS to automatically search for the optimal cross-attention network architecture and corresponding interaction functions. Finally, we integrate LLMs with the aligned user-item features to generate personalized explanations that reflect individual user preferences and item characteristics. Extensive experiments on real-world datasets demonstrate the effectiveness of our proposed approach.

We summarize our contributions as follows:
\begin{itemize}
    \item {\bfseries Problem Formulation:} We formulate the problem of effective explanation generation in recommender systems and identify the key challenges inspired by the real applications. 
    \item {\bfseries Model:} We propose an end-to-end Bi-level Neural Architecture Search (NAS) framework that automatically searches for the optimal explainable architecture for generating effective explanations while enhancing personalization through zero-shot prompting strategies.
    \item {\bfseries Comprehensive Evaluation:}  We conduct extensive experiments on four real-world datasets, evaluating both recommendation performance and explanation quality. The results demonstrate the superior performance of our approach across multiple evaluation metrics.
\end{itemize}

\begin{figure*}[!thb]
\centering
\includegraphics[width=0.9\textwidth]{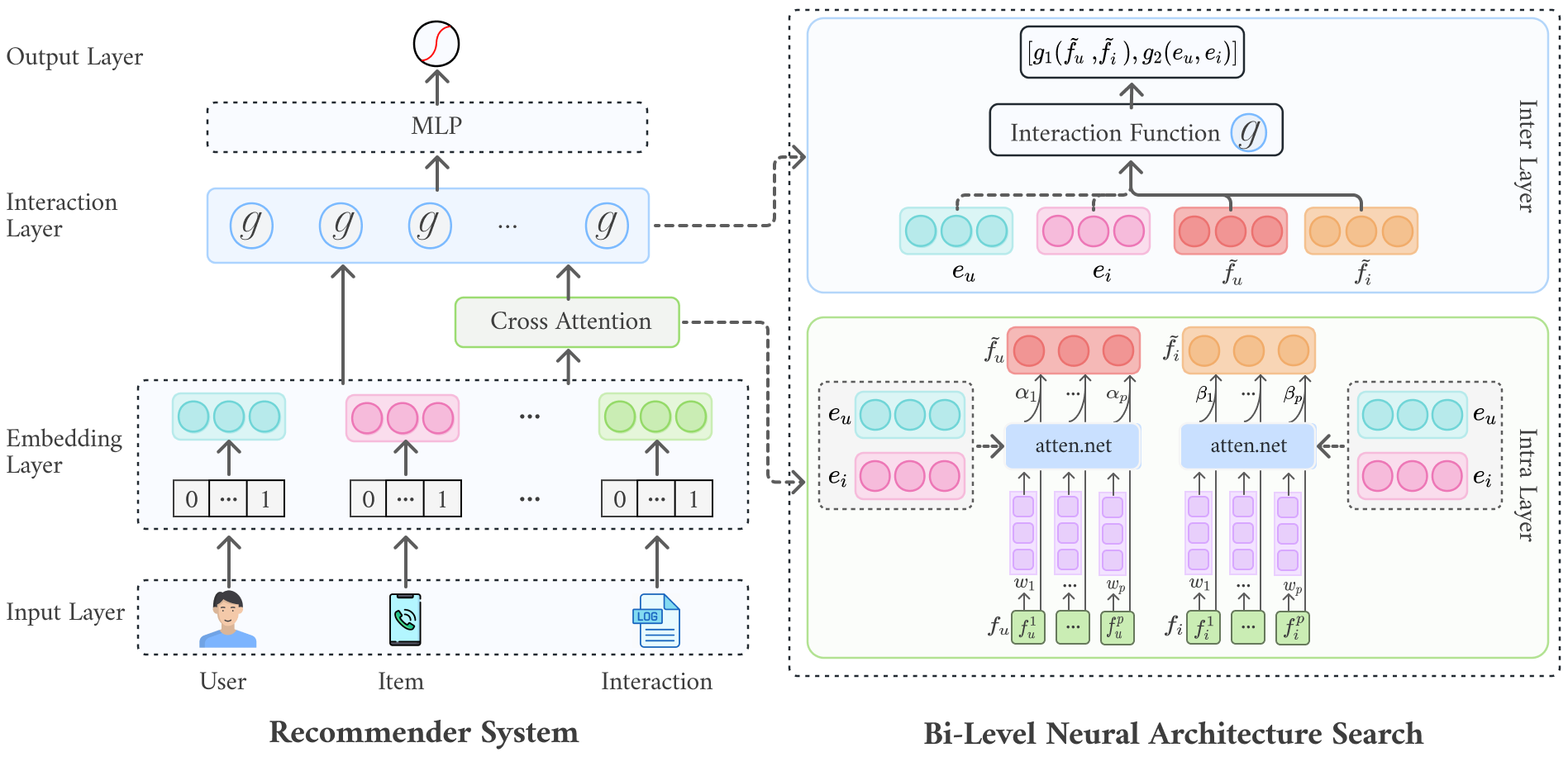}
\vspace{-2mm}
\caption{Overview of our proposed architecture. The left part is the architecture for recommender systems, and the right part illustrates the framework of bi-level NAS, which can jointly discover the optimal cross-attention construction and feature interaction functions.}
\label{fig: general architecture for RS}
\vspace{-3mm}
\end{figure*}

\section{Problem Definition}
In this section, we introduce the preliminaries of our problem setting. Then, we give the formal definition of explainable recommendation with NAS. We use upper case calligraphic font letters to denote sets (e.g., $\mathcal{U}$), bold upper case letters to denote matrices (e.g., $\bm{A}$), bold lower case letters to denote vectors (e.g., $\bm{x}$), and regular lower case letters to denote scalars (e.g., $\alpha$). 

\textbf{Preliminaries and Notations. } 
Assume we have a user set with $M$ users $\mathcal{U}=\{u_1,u_2,\cdots,u_M\}$ , an item set with $N$ items $\mathcal{I}=\{i_1,i_2,\cdots,i_N\}$, and a feature set with $P$ features $\mathcal{F}=\{f_1, f_2, \cdots,f_P\}$. 
We apply the sentiment analysis toolkit ~\cite{zhang2014users} to extract (Aspect, Opinion, Sentiment) triplets from the textual reviews. For example, in the domain of clothing, the extracted features (aspects) include price, quality, color, size, etc. The corresponding opinions include high, good, great, fit, etc., and the sentiment could be either +1 or -1. 
Following the similar pre-processing steps adopted by previous work ~\cite{zhang2014explicit, wang2018explainable, chen2020try, tan2021counterfactual, zhou2021intrinsic},
we compute the user-feature preference vector $f_{u} \in \mathbb{R}^{p}$ and $u \in \mathcal{U}$, and the item-feature quality vector $f_{i} \in \mathbb{R}^{p}$ and $i \in \mathcal{I}$. $f_{u}^k$ indicates to what extent the user $u$ cares about the feature $f_k$. Considering that user tends to comment on the features they care more about and may express their opinions more frequently on those features, it is a frequency-based vector. $f_{i}^k$ indicates how well the item $i$ performs on the feature $f_k$. Considering that the quality of items can be reflected in the review sentiments from users, it is formulated by considering both review feature frequency and average sentiment. Specifically,
\begin{equation}
\label{eqn: review feature score}
\begin{split}
f_{u}^k =&\begin{cases} 
            0,  \text{if user $u$ did not mention feature $f_k$} \\
            1+(T-1)(\frac{2}{1+\mathrm{exp}(-c_{u}^k)}-1),   \text{otherwise}
         \end{cases}
         \\
f_{i}^k =& \begin{cases} 
            0,  \text{if item $i$ was not mentioned with feature $f_k$} \\
            1+(\frac{T-1}{1+\mathrm{exp}(-c_{i}^k \cdot s_{i}^k)}),   \text{otherwise}
         \end{cases}
\end{split}
\end{equation}
where $T$ is the maximum review score,
$c_{u}^k$ is the frequency that user $u$ mentioned feature $f_k$, $c_{i}^k$ is the frequency that item $i$ is mentioned with feature $f_k$, while $s_{i}^k$ is the average sentiment of item $i$ regarding feature $f_k$. 
After this transformation, all entries of these feature vectors are mapped into the range of $[0, T]$.

\textbf{Cross-Attention.} 
Different from self-attention, which operates on a single embedding vector, cross-attention typically involves two distinct vectors of the same dimension. It calculates attention scores for each dimension in one vector based on its relevance to another vector and integrates the two vectors. 
In this paper, we utilize cross-attention between user/item embeddings and their corresponding feature embeddings to explore their interconnections. 



\textbf{Neural Architecture Search.} 
Neural Architecture Search refers to the systematic process of automating the design of neural network architectures. It includes three main parts: search space, search strategy, and performance evaluation strategy.
Designing a search space is a crucial aspect of NAS as it directly impacts the efficiency and effectiveness of the entire process~\cite{ren2021comprehensive}. The search space determines the realm of architectural possibilities, allowing NAS to explore diverse neural network configurations.
On one hand, the space needs to be broad enough to capture a diverse range of architectural possibilities, allowing for innovation and inclusion of human knowledge. On the other hand, the space cannot be too general, as that would lead to computational inefficiency and high exploration costs~\cite{yao2020efficient}.
The search strategy (optimization methods) details how to explore the search space. It involves the classical trade-off between exploration and exploitation. On the one hand, there is a desire to rapidly discover high-performing architectures. On the other hand, it's crucial to avoid early convergence with suboptimal architectures ~\cite{wistuba2019survey}. 
The evaluation process assesses the generalization performance of each architecture on unseen data, offering insights into its effectiveness and aiding in the identification of optimal architectures~\cite{zoph2018learning}.

In this paper, we delve into the realm of recommendation systems, focusing on the scenario of providing effective explanations. For each user-item pair $(u, v)$, we deduce the preferences of user $u$ and features of item $v$ from reviews. An effective explainable recommendation system should recommend items with features that closely align with users' preferences and provide the corresponding common features for the explanation. Here, we give the formal definition of our recommendation problem:

{\setlength{\parindent}{0pt}
\begin{problem}
 \textbf{Effective Explainable Recommendation} \\
    \textbf{Input:} (i) A set of users $\mathcal{U} = \{u_1, u_2,\cdots, u_M\}$, (ii) A set of items $\mathcal{I} = \{i_1, i_2,\cdots, i_N\}$, (iii) A set of raw features $\mathcal{F} = \{f_1, f_2,\cdots, f_P\}$, and (iv) the observed user-item-interaction sets $\Omega$. \\
   \textbf{Output:} The estimated interaction scores of the unobserved items for each user $u \in \mathcal{U}$ as well as their corresponding feature explanations.
   \end{problem}
}
\section{Methodology}

In this section, we go over the details of our proposed framework, including the recommendation component, the NAS-driven architecture search space, and our optimization strategy. We begin by using NAS to automatically explore a predefined architecture space, optimizing performance under resource constraints with minimal human intervention and tailoring architectures to each dataset. Next, we leverage LLMs to generate personalized, context-aware explanations from the NAS-aligned user–item features, enriched by user history. Finally, we present illustrative examples of our results.

\subsection{Recommendation Component}
Existing deep recommender system frameworks usually share a similar structure, as shown in Figure~\ref{fig: general architecture for RS}. They have three modules: embedding layer, interaction layer, and output layer~\cite{zhao2021autoloss,lin2022adafs}. We provide a high-level overview in this subsection.

\textbf{Embedding Layer.} 
The input of the embedding layer includes both user and item IDs as well as various user and item features, and the output is the embeddings for each of these fields. 
In this paper, we focus on using features extracted from reviews for both users and items. We embed each word within these features into a lower-dimensional representation using pre-trained word embeddings (e.g., GloVe).
Given the high dimensionality and sparsity of one-hot encoding for user and item IDs, as well as the difference in their corresponding embedding size, we also introduced an embedding layer to transform the user and item IDs into a low-dimensional continuous vector, the final output of the embedding layer is:
\begin{equation}
\label{eqn:embedding layer}
\begin{split}
E=[e_u, e_i, f_u, f_i]
\end{split}
\end{equation}
where $e_u$ and $e_i$ are the embeddings for user and item IDs, while $f_u$ and $f_i$ are the embeddings of their corresponding features.

\textbf{Interaction Layer.} 
The interaction layer receives embeddings from the embedding layer and aims to capture the interactions between features explicitly. 
This layer can be easily extended to capture the higher order of feature interaction. The most commonly used way is to adopt element-wise operations to capture second-order feature interaction. 
For simplicity, we adopt the element-wise operations without the loss of generality to capture pairwise interactions.
\begin{equation}
\label{eqn:user_item_interaction}
\begin{split}
I=g(f_u, f_i)
\end{split}
\end{equation}
where $g(\cdot)$ represents the interaction function between features, and $f_u$ and $f_i$ are embeddings of features from user $u$ and item $i$. The function $g$ may take various forms, such as inner product, plus/minus, max/min, and so forth.

\textbf{Output Layer.} 
The output layer learns the transformation and combination of features, i.e., $E$ and $I$, and outputs the probability score $\hat{y}$ for the interaction between specific users and items. Typically, it consists of several fully connected layers with non-linear activation functions. 
The selection of activation function $\mathrm{\sigma(\cdot)}$ depends on the specific recommendation task.
In this paper, sigmoid has been chosen as the activation function.

\subsection{Architecture Search Space}
In this subsection, we define a search space that considers intra-layer design and inter-layer design. We propose first searching for a way of constructing cross-attention and then integrating these vectors. Specifically, 
\begin{itemize}
\item {\bfseries Intra-layer}: A possible way to construct cross-attention.
\item {\bfseries Inter-layer}: Interaction functions that operate on users/items and their corresponding features.
\end{itemize}

\textbf{Intra-layer Design.} 
Intuitively, in a specific application domain, users tend to concentrate on particular features that matter most to them. Similarly, within this context, each item is expected to have a distinct set of features that are particularly attractive to its potential customers. This selective focus on relevant features by both users and items is a fundamental aspect of explainable recommender systems, as it allows for the delivery of highly customized and meaningful recommendations.
Inspired by this observation, we highlight the significance of these essential features by designing the attention network as: 
\begin{equation}
\label{eqn:intra-layer design}
\begin{split}
\alpha_k=\frac{\mathrm{exp}(e_uM_aw_k)}{\sum_{k=1}^{p}\mathrm{exp}(e_uM_aw_k)}
\end{split}
\end{equation}
where $e_u \in \mathbb{R}^{m}$ and $w_k \in \mathbb{R}^{p}$ are the embedding of user $u$ and $k$-th feature of user review, respectively. $M_a \in \mathbb{R}^{m\times p}$ denotes the attention matrix between user embedding and feature embedding. The attention weight for each user towards $k$-th feature is represented as $\alpha_k \in (0,1)$. Likewise, each item can also obtain its own attention weight denoted as $\beta_k \in (0,1)$ with a different attention mapping matrix $M_b \in \mathbb{R}^{m\times p}$. 
Then, the attention-aggregated representations of users/items features can be obtained as follows:
\begin{equation}
\label{eqn:tilda}
\begin{split}
\tilde{f_u} = [\alpha_1,\cdots,\alpha_p] \odot f_u \quad \text{and} \quad
\tilde{f_i} = [\beta_1,\cdots,\beta_p] \odot f_i 
\end{split}
\end{equation}

As shown in Equation~\ref{eqn:intra-layer design}, the attention matrix $M_a$ for user features is learned from user embedding, and the attention matrix $M_b$ for item features is learned from item embedding, respectively, $\mathrm{att}_{0}$ denotes this way of cross-attention construction. However, user preference and item characteristics often share a complex interplay: user preference may exert influence on both user and item features (denoted as $\mathrm{att}_{1}$), while item characteristics can also, in turn, shape user preferences and item features (denoted as $\mathrm{att}_{2}$). This dynamic interaction can also manifest in a reciprocal manner, where users and items cross-influence each other (denoted as $\mathrm{att}_{3}$). To encapsulate these multifaceted relationships, we introduce four interconnected constructions of cross-attention ($\mathrm{att}_{0-3}$), which is depicted in the lower right corner of Figure~\ref{fig: general architecture for RS}.

\textbf{Inter-layer Design.} 
Having defined the intra-layer for constructing cross-attention, the next crucial aspect of design involves the fusion of these results. As illustrated in Figure~\ref{fig: general architecture for RS}, the input of the feature interaction layer includes embeddings of users and items, as well as the output obtained from the cross-attention. In this context, the operation set is composed of three commonly employed element-wise interaction functions: $O = \{\mathrm{Plus}, \mathrm{Multiply}, \mathrm{Concat}\}$. Each of these interaction functions carries its unique characteristics. The `Plus' operator encourages a cooperative interaction, emphasizing the additive combination of its inputs. The `Multiply' operator fosters a more selective and focused interaction, highlighting the mutual influence between its inputs, while the `Concat' operator expands the feature space, allowing a more complex and expressive representation of their interaction. 
More specifically, 
\begin{equation}
\label{eqn:inter layer design}
\begin{split}
I=[g_1(\tilde{f_u}, \tilde{f_i}), g_2(e_u,e_i)]
\end{split}
\end{equation}
where the operators $g_1, g_2 \in O$.

It's worth noting that the operators need not necessarily be the same, enabling us the flexibility to employ diverse interaction strategies. Moreover, to address the potential discrepancy in the embedding dimensions between users/items and the features associated with users/items, we explicitly utilize the `Concat' operator, which seamlessly merges these distinct components, addressing dimension-related issues. Additionally, the second part $g_2$  may not always be a mandatory element in our design. We consider the option of keeping one of them ($e_i$ or $e_u$) or removing both of them. 

\begin{table}[htp]
\caption{Designed search space}
\centering
\label{tab: search space}
 \begin{tabular}{cccc}
\hline
\textbf{\# of Inputs}& \textbf{Inputs}& \textbf{Intra-layer}& \textbf{Inter-layer}\\
\hline 
{2} & $\tilde{f_u}, \tilde{f_i}$  & $\mathrm{att} \in \mathrm{att}_{0-3}$  &$g_1 \in O$  \\
\hline
\multirow{2}{*}{3}
        &$\tilde{f_u}, \tilde{f_i}, e_u$  & $\mathrm{att} \in \mathrm{att}_{0-3}$  & $g_1 \in O$ \\
        &$\tilde{f_u}, \tilde{f_i}, e_i$  & $\mathrm{att} \in \mathrm{att}_{0-3}$  & $g_1 \in O$ \\
\hline
{4} &$\tilde{f_u}, \tilde{f_i}, e_u, e_i$  & $\mathrm{att} \in \mathrm{att}_{0-3}$ & $g_1, g_2 \in O$ \\
\hline
\end{tabular}
\end{table}

As shown in Table~\ref{tab: search space}, the scale of the search space is determined by both the intra-layer design and inter-layer design. 
Precisely, the size of the search space for the intra-layer is $|\mathrm{att}_{0-3}|=4$, and the size of the search space for the inter-layer can be computed as $3+2*3+3*3=18$. Thus, the possible number of architectures in the search space is $18*4=72$.

The intra-layer and inter-layer design equip our recommender system with the versatility needed to adapt and optimize its performance across diverse recommendation scenarios. Through a careful selection and combination of these functions, we can finetune the system to provide personalized and effective recommendations that cater to the 
preferences and interests of every user.

\begin{figure}
    \centering
    \includegraphics[width=.5\textwidth]{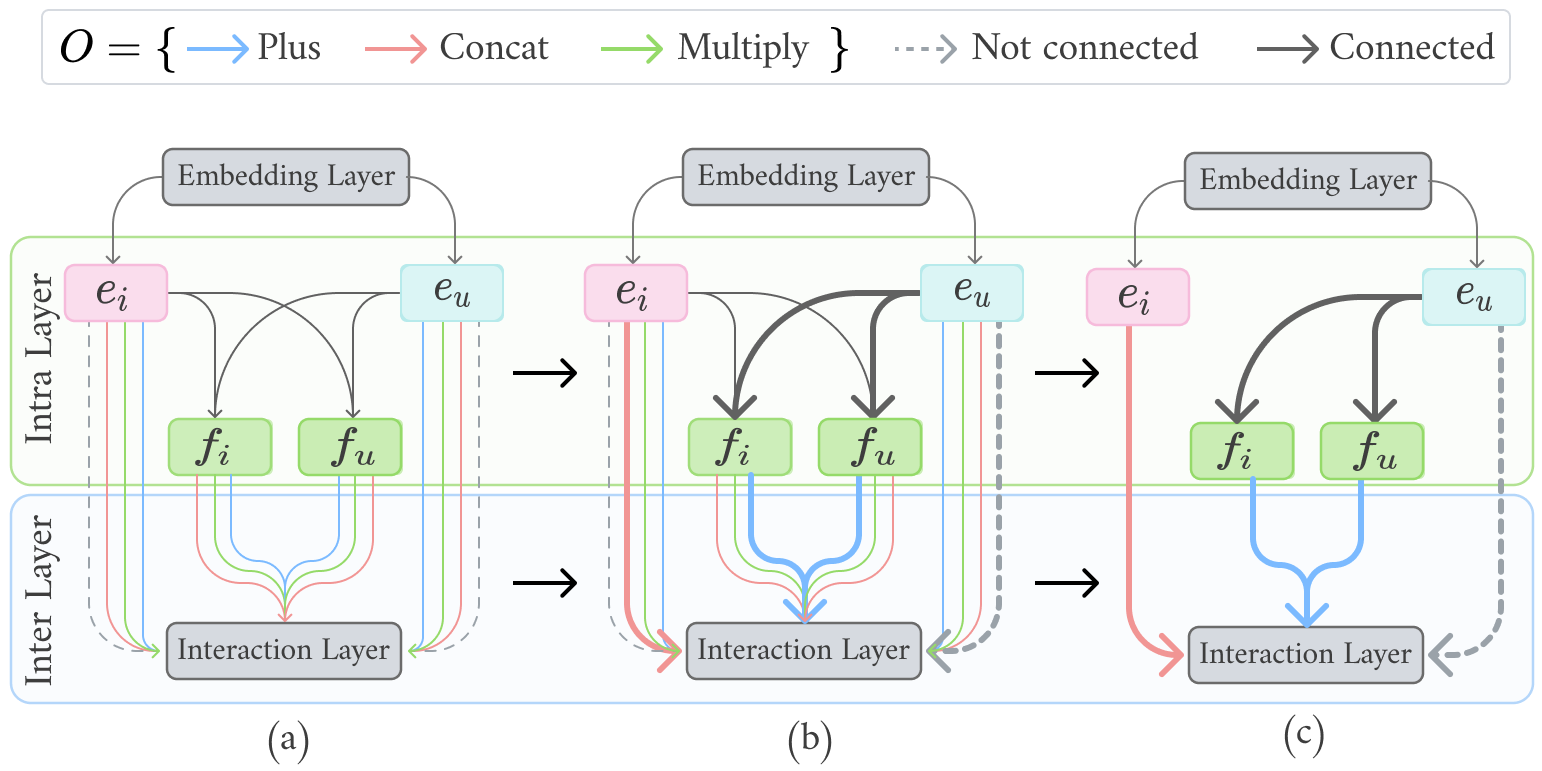}
    \caption{An overview of the optimization strategy. Each node denotes the latent representation to be learned, and each edge denotes an operation (e.g., plus, multiply, etc., indicated by color). (a) Relax the choices of operation in a continuous space by placing a mixture of all candidate operations on each edge. (b) Utilize the bi-level NAS to learn the weights of each edge. (c) Finalize the architecture from the edges with the largest weights.}
    \label{fig:search space} 
\end{figure}
\subsection{Optimization Strategy}
Generally speaking, the major challenge in NAS lies in its substantial computational cost of training and evaluating numerous neural network architectures. To address this issue, we introduce Bi-level NAS (Bi-NAS), a method that efficiently conducts architecture searches within the predefined search space by jointly optimizing intra-layer and inter-layer configurations in an end-to-end manner.

As depicted in Figure~\ref{fig:search space}, our proposed architecture can be viewed as a directed acyclic graph (DAG) comprising sequential nodes. Each node denotes the latent representation to be learned (e.g., the embedding of users), and each directed edge denotes an operation (e.g., plus) that transforms the corresponding node. The value of each intermediate node is calculated based on all of its preceding nodes. Additionally, we introduce a special ``Not connected" operation to indicate when there's no connection between two nodes.

It is worth noting that the search space includes both discrete elements(e.g., selection of operation $\bm{\lambda}$) and continuous variables (e.g., hyper-parameters $\bm{w}$). Conducting an exhaustive search without considering the diverse search space is inherently inefficient. Inspired by recent advances in the differentiable search of NAS, we propose to relax the choices among operations as a sparse vector in a continuous space (shown in Figure~\ref{fig:search space}(a)). The main idea is to parameterize the architecture using continuous variables, making it differentiable with respect to these parameters. Thus, the gradient descent could be utilized to optimize the architecture directly. Specifically, let $\bm{\lambda} = [\lambda_t] \in \mathbb{R}^{|O|}$ where $\lambda_t$ denotes the weight of the $t$-th operation. The selection of a specific operation from a categorical choice can be relaxed to a mixture of all possible operations:
\begin{equation}
\label{eqn: operation selection}
\begin{split}
\hat{g}=\sum_{t=1}^{|O|}\lambda_t(g_t(\tilde{f_u},\tilde{f_i}))), \: s.t. \: \bm{\lambda} \in \ \mathcal{C}
\end{split}
\end{equation}
where $g_t \in O$ is the $t$-th operation of $O$, and the constraint set 
$\mathcal{C}=\{\bm{\lambda}, ||\bm{\lambda}||_0=1 $ and $ 0 \leq \lambda_t \leq 1\}$. 

At the end of searching, a discrete operation can be obtained by replacing mixed operations with the most likely operation with the largest $\lambda$, i.e., $g=\argmax_{g_i \in O} \lambda_i$ (shown in Figure~\ref{fig:search space}(c)).
This constraint guarantees that only one operation can be chosen, and the model's architecture can be reflected by $\bm{\lambda}$. The task of operation search then reduces to learning a set of continuous variables $\bm{\lambda}$. Our goal is to jointly learn the architecture $\bm{\lambda}$ and the associated weights $w$,
striving for a unified system that integrates architectural design and parameter selection to improve performance and efficiency.

We optimize the following bi-level objective function:


\begin{equation}
\label{eqn:loss}
\begin{split}
\mathcal{G}
=\argmin_{\bm{\lambda}} \mathcal{L}_{\text{val}}(\hat{w}(\lambda), \lambda) \\
\text{s.t.} \quad \hat{w} =\argmin_{\bm{w}} \mathcal{L}_{\text{train}}(w, \lambda) \\
\end{split}
\end{equation}

where $\mathcal{L}_{\text{train}}$ and $\mathcal{L}_{\text{val}}$ are the loss on the training set and validation set. Specifically, $\mathcal{L}=\mathcal{L}(y, \text{MLP}_{w}(\mathrm{att}_i, \hat{g}))$ is the binary cross-entropy loss, $y$ is the ground truth label. And $y=1$ indicates that user $u$ interacted with item $i$, otherwise $y=0$. $\mathrm{att}_i \in \mathrm{att}_{0-3}$, $\mathrm{MLP}(\cdot)$ is the output of output layer.

\begin{figure}[h]
    \centering
    \includegraphics[width=.48\textwidth]{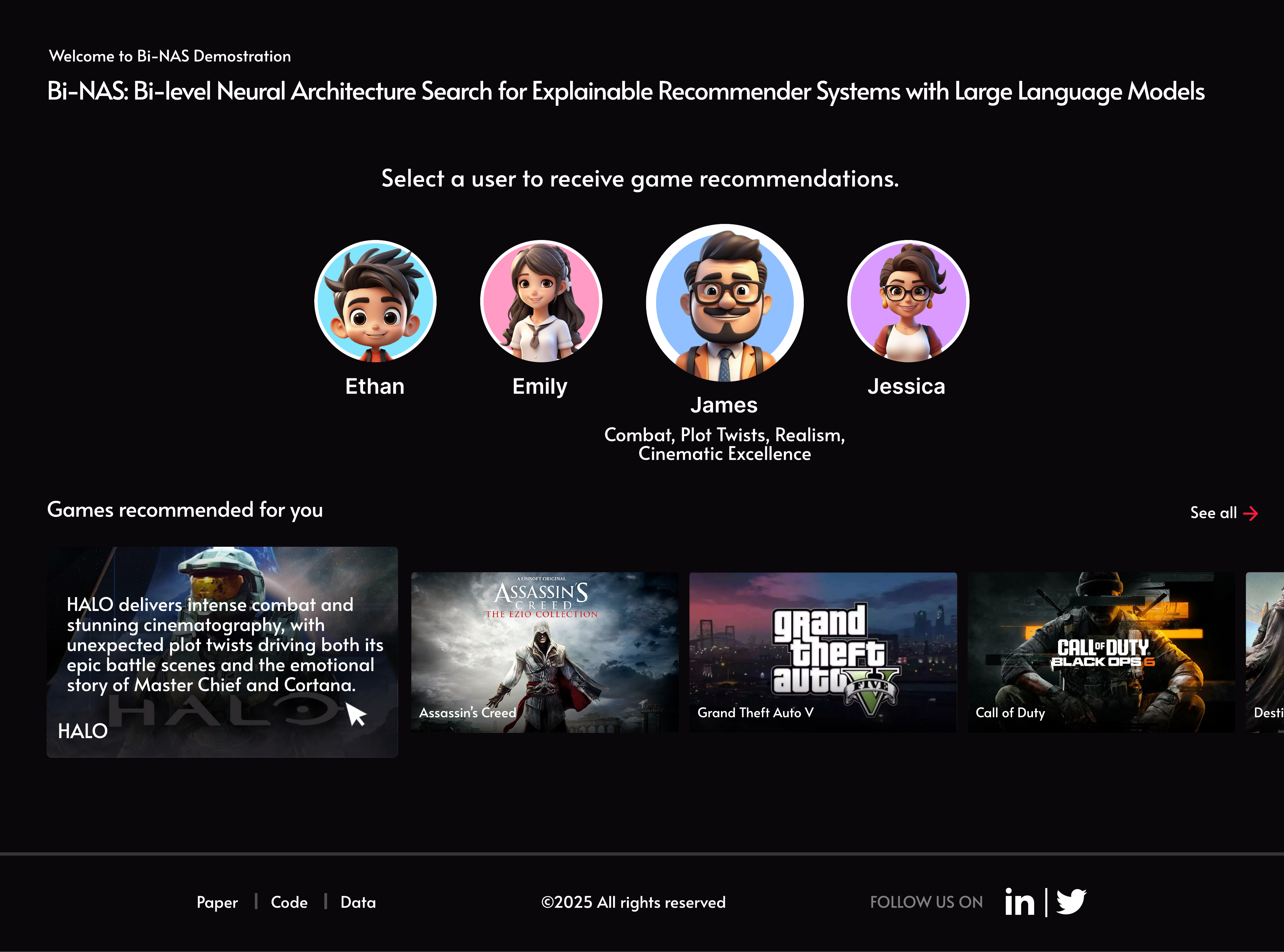}
    \caption{A demonstration of the generated explanations for the recommendation list.}
    \label{fig: demo}
    \vspace{-3mm}
\end{figure}

\begin{table*}[!t]
    \caption{{Recommendation performance evaluation on four datasets. The numbers in bold represent the best results, while the numbers with a wave represent the second-best results.}} 
    \label{tab: recommendation performance}
    \centering
    \scalebox{0.85}{
    \begin{tabular}{c ccc ccc ccc ccc}
        \toprule
        \multirow{2}{*}{\textbf{Method}} &
        \multicolumn{3}{c}{\textbf{Video}} &
        \multicolumn{3}{c}{\textbf{Instrument}} &
        \multicolumn{3}{c}{\textbf{Beauty}} &
        \multicolumn{3}{c}{\textbf{Clothing}} \\
        \cmidrule(lr){2-4} \cmidrule(lr){5-7} \cmidrule(lr){8-10} \cmidrule(lr){11-13}
        & Hit@10 & NDCG@10 & MRR & Hit@10 & NDCG@10 & MRR & Hit@10 & NDCG@10 & MRR & Hit@10 & NDCG@10 & MRR\\
        \midrule
        NCF   & 0.505 & \uwave{0.295}  & 0.230 & \uwave{0.300} & \uwave{0.172}  & \uwave{0.132}  & 0.413  & 0.251 & 0.202 & 0.317 & 0.178 & 0.136 \\
        VBPR  & 0.507 & 0.285  & 0.224 & 0.262 & 0.140  & 0.108  & 0.401 & 0.225 & 0.191 & 0.289 & 0.155 & 0.123\\
        CER  & 0.463 & 0.267  & 0.207 & 0.259 & 0.145  & 0.111  & \textbf{0.559}  & 0.300 & 0.221 & 0.396 & 0.220 & 0.167\\
        NAR   & \uwave{0.520} & 0.294  & \uwave{0.231} & 0.284 & 0.151  & 0.117  & 0.506  & 0.297 & \uwave{0.256} & \uwave{0.400} & \uwave{0.221} & \uwave{0.178} \\
        MANAS  & 0.341 & 0.202  & 0.176 & 0.148 & 0.072  & 0.067  & 0.477  & \uwave{0.315} & \textbf{0.278} & 0.204 & 0.109 & 0.098\\
        \cmidrule(lr){1-1} \cmidrule(lr){2-4} \cmidrule(lr){5-7} \cmidrule(lr){8-10} \cmidrule(lr){11-13}
        \textbf{Ours}  & \textbf{0.543} & \textbf{0.307}  & \textbf{0.241} & \textbf{0.342}  & \textbf{0.190} & \textbf{0.150}  & \uwave{0.545} & \textbf{0.322} & \textbf{0.278} & \textbf{0.406} & \textbf{0.227}  & \textbf{0.185}\\
        \bottomrule
    \end{tabular}
    }
\end{table*}

\subsection{Effective and Personalized Explanation Generation}

\begin{figure}[h]
    \centering
    \includegraphics[width=.48\textwidth]{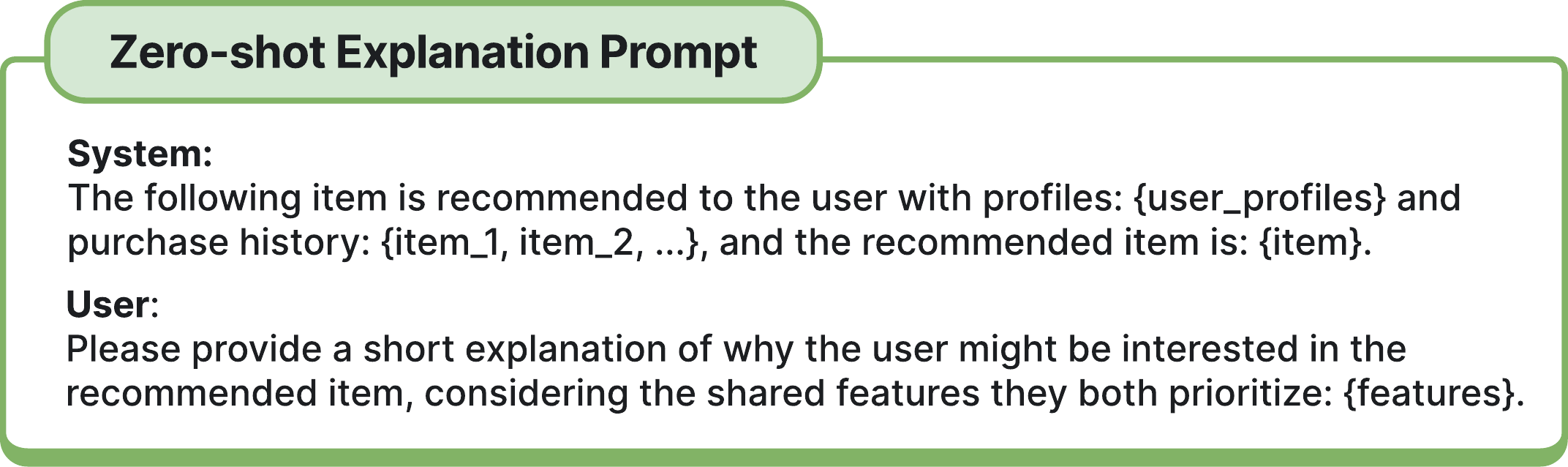}
    \caption{Zero-shot explanation prompt template for enhancing effective and personalized explanation generation.}
    \label{fig:explanation prompt}
\end{figure}

To illustrate the explanations generated by our framework, we present some examples through an integrated user interface. For this demonstration, we utilize data from the Amazon-Video dataset~\cite{ni2019justifying} and employ Llama-3.1-8B-Instruct~\cite{dubey2024llama} as our primary LLM. 
First, we generate a general user profile based on the features they prioritize. Then, we recommended the five most relevant items based on the user's history. For each recommended item, we leveraged the LLM to generate an explanation for why it was suggested.
Figure~\ref{fig: demo} illustrates the interactive user interface used in our demonstration. The interface consists of two main sections: the top section displays multiple user accounts logged into the device, each with their respective preferences represented. For example, a dad's profile may prioritize video games featuring combat, plot twists, realism, and cinematic excellence, while a mom's profile may favor games centered on character development, storytelling, drama, and engaging scripts. Once an account is selected, the bottom section presents the personalized recommendations. Users can hover over each recommended item to view its detailed explanation. Unlike traditional methods that generate explanations directly using LLMs, our approach primarily derives explanations from the aligned features shared between the user and the recommended item and user history, as outlined in Figure~\ref{fig:explanation prompt}. For example, the explanation why Halo is recommended for the role of dad is because: Halo delivers intense combat and stunning cinematography, with unexpected plot twists driving both its epic battle scenes and the emotional story of Master Chief and Cortana.

\section{Experiments}
In this section, we assess our method's performance and explanation quality, conduct an ablation study on cross-attention effectiveness, and analyze the obtained architecture.

\subsection{Experimental Settings}
\textbf{Dataset.}
We utilize the Amazon dataset~\cite{ni2019justifying}, where the raw data is collected as a user-item-review triplet format. We conduct experiments on its top-level product categories\footnote{\url{https://jmcauley.ucsd.edu/data/amazon/}}: Instrument, Video, Beauty, and Clothing. 
\begin{table}[htp]
\caption{Statistics of the datasets. The density is computed as $\# \: \text{Interactions}/(\# \: \text{Users} \times \# \: \text{Items})$.}
\centering
\label{tab: statistics}
\scalebox{0.78}{
\begin{tabular}{cccccc}
\hline
\textbf{Datasets}& \# \textbf{Users}& \# \textbf{Items}& \# \textbf{Interactions}& \# \textbf{Features}& \textbf{Density}\\
\hline 
Instrument &1,276  & 843  &3,581   &325 & 0.33\%\\
Video &4,333  &1,486  &11,759   &350 & 0.18\%\\
Beauty &21,472  & 11,897  &105,659   &1,985 & 0.04\%\\
Clothing &37,703  &22,647  &142,553   &1,462 & 0.02\%\\
\hline
\end{tabular}}
\end{table}

The statistics of these datasets are summarized in Table~\ref{tab: statistics}. In preprocessing, we keep the user-item interactions that have text reviews. We treat the 4-star and 5-star reviews as positive feedback, and the rest are unlabeled feedback where we performed negative sampling within it during training. Following this preprocessing protocol, all user and item interactions are eventually stored in the interaction matrix with binary values $0$ and $1$. Additionally, we utilize Sentire\footnote{\url{https://github.com/lileipisces/Sentires-Guide}} to get phrase-level aspect words and sentiment.

\noindent\textbf{Evaluation protocols.}
We randomly split the dataset into 70\%, 20\%, and 10\% for training, validation, and testing, respectively, considering only users with more than five positive samples.
 We evaluate the performance of all methods by three standard
measurements for ranking tasks: Hit Rate (Hit@10), Normalized Discounted Cumulative Gain (NDCG@10), and Mean Reciprocal Rank (MRR). 


\begin{table*}[h]
    \caption{Explanation evaluation on four datasets. The numbers in bold represent the best results, while the numbers with a wave represent the second-best results.}
    \label{tab: explanation performance-short}
    \centering
    \scalebox{0.95}
    {
    \begin{tabular}{ccccccccccccccccc}
        \hline
        \multirow{2}{*}{\textbf{Method}} &
        \multicolumn{4}{c}{\textbf{Video}} &
        \multicolumn{4}{c}{\textbf{Instrument}} &
        \multicolumn{4}{c}{\textbf{Beauty}} &
        \multicolumn{4}{c}{\textbf{Clothing}} \\
        \hhline{~~---~---~---~---}
        & 
        & Pre & Recall & $F_1$  &
        & Pre & Recall & $F_1$  &
        & Pre & Recall & $F_1$  &
        & Pre & Recall & $F_1$  \\
        \hline
        Random &
        & 0.007 & 0.027 & 0.011 &
        & 0.010 & 0.032 & 0.014 &
        & 0.019 & 0.028 & 0.019 &
        & 0.015 & 0.028 & 0.018 \\
        CER &
        & 0.100 & 0.448 & 0.154 & 
        & 0.098 & 0.383 & 0.146 & 
        & 0.199 & 0.358 & 0.225 & 
        & 0.161 & 0.347 & 0.201 \\
        NAR &
        & \uwave{0.119} & \uwave{0.452} & \uwave{0.174} &
        & \uwave{0.135} & \uwave{0.442} & \uwave{0.192} &
        & \uwave{0.233} & \uwave{0.385} & \uwave{0.254} &
        & \uwave{0.195} & \uwave{0.381} & \uwave{0.235} \\
        \hhline{-~---~---~---~---}
        \textbf{Ours} &
        &\textbf{0.137} & \textbf{0.528} & \textbf{0.201}  &
        &\textbf{0.143} & \textbf{0.451} & \textbf{0.201} &
        &\textbf{0.239} & \textbf{0.403} & \textbf{0.262} &
        &\textbf{0.206} & \textbf{0.395} & \textbf{0.247}\\
        \hline
    \end{tabular}
    }
\end{table*}

\subsection{Baselines}
We conduct comparison experiments between our method and the following state-of-the-art approaches. We follow the official implementation and use the default setting on hyperparameters for these baselines. 
\textbf{NCF}~\cite{he2017neural}: One of the most popular embedding-based neural network recommendation methods. It replaces the inner product from matrix factorization with a neural architecture that can learn an arbitrary function from data.
\textbf{VBPR}~\cite{he2016vbpr}: A modified version of the original VBPR by using the neural network-based embeddings and the contextual features.
\textbf{CER}~\cite{tan2021counterfactual}: A counterfactual explainable recommendation model. This model leverages counterfactual reasoning from causal inference to deliver explainable recommendations, aiming to provide simple and effective explanations for the model decision. 
\textbf{NAR}~\cite{zhou2024based}: The attention-based intrinsic recommendation model, has intrinsically explainable designs with two major components: a representation network for users and items, and a prediction network to predict the relevance scores. Additionally, its model architecture can be searched using our Bi-NAS framework.
\textbf{MANAS}~\cite{chen2022learn}: MANAS automatically assembles the three basic logical operation modules into a network architecture tailored to the given user by neural architecture search.

\subsection{Performance Evaluation}
We first evaluate the overall performance of the recommendation task of our proposed method. The results are reported in Table~\ref{tab: recommendation performance}.
In general, we have the following observations: 
(1) Our proposed method can outperform all other baselines in most of the cases in terms of various evaluation metrics, regardless of the dataset size.
(2) Furthermore, other models like CER and NAR also exhibit competitive performance. Specifically, CER shows competitive performance when it comes to large-scale datasets; we believe this is due to the factor that CER is more reliant on reviews, which tend to be richer and more informative in larger datasets. 
(3) In contrast, MANAS does not perform as well as other baseline methods. We believe this disparity in performance is partly due to this method's reliance on sequential information without leveraging review data.
(4) Our model consistently outperforms NAR, an attention-based model with a fixed cross-attention and interaction configuration. The fixed architecture does not generalize well across different dataset, which could further demonstrate the effectiveness of our method.

\subsection{Explanation Evaluation}

In this subsection, we want to evaluate the explainability based on the user's review of the item, which has been widely used in previous work ~\cite{abdollahi2016explainable, chen2019co, tan2021counterfactual}. Specifically, we utilize the review as the ground truth to explain why the item is recommended to the user. 
We measure the performance of the explanation by metrics utilized in other work~\cite{tan2021counterfactual, zhou2021intrinsic}. Here, $R_{u,i}^{k}$ denotes the interactions between user $u$ and item $i$ regrading feature $f_k$. If the review between user $u$ and item $i$ has positive sentiment on the feature $f_k$, then $R_{u,i}^{k}=1$; otherwise, $R_{u,i}^{k}=0$.  
\begin{equation}
\label{eqn: pre_recall}
\begin{split}
\mathrm{Precision} = \frac{\sum_{k=1}^{p}R_{u,i}^{k}\cdot I_{u,i}^{k}}{\sum_{k=1}^{p} I_{u,i}^{k}}, 
\mathrm{Recall} = \frac{\sum_{k=1}^{p}R_{u,i}^{k}\cdot I_{u,i}^{k}}{\sum_{k=1}^{p}R_{u,i}^{k}}
\end{split}
\end{equation}
where $I_{i,j}^{k}$ is an identity function to identify the feature explanation obtained by our model. If the obtained explanation identify item $i$ is recommended to user $u$ because of feature $f_k$, then $I_{u,i}^{k}=1$; otherwise $I_{u,i}^{k}=0$.
Precision is defined as the ratio of features that users really like in the obtained feature explanation, Recall is defined as the ratio of features explanation to the number of features that users like, and F1 is represented as the harmonic mean of Precision and Recall. NDCG can be computed similarly.  

Furthermore, we randomly choose features from the feature space to generate explanations. The results of this random baseline can provide some insight into the difficulty of this task.
The results are reported in Table~\ref{tab: explanation performance-short}. In general, we have the following observations:
(1) Our method consistently performs the best in terms of all metrics, demonstrating that our method can provide effective explanations.
(2) CER does not perform as well as other methods. We believe this is because CER employs item-oriented perturbations for explanations. Consequently, when explaining a user's ranking decision, the perturbations from all the items they've reviewed are averaged, leading to a reduction in granularity for explanation.
(3) Although NAR has an explicitly designed architecture for learning the attention weight, it remains outperformed by our method. This implies that human prior knowledge can occasionally constrain the potential of models.

\begin{figure}[h]
\small
\centering
\subfigure[Performance evaluation]{
    \includegraphics[width=.225\textwidth]{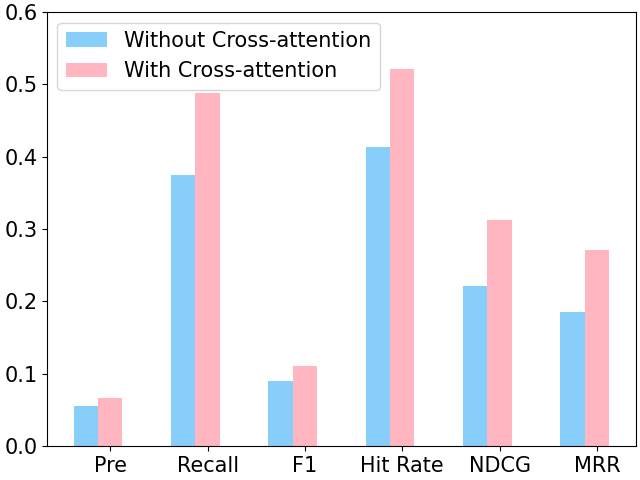}
    }
\subfigure[Explanation evaluation]{
    \includegraphics[width=.225\textwidth]{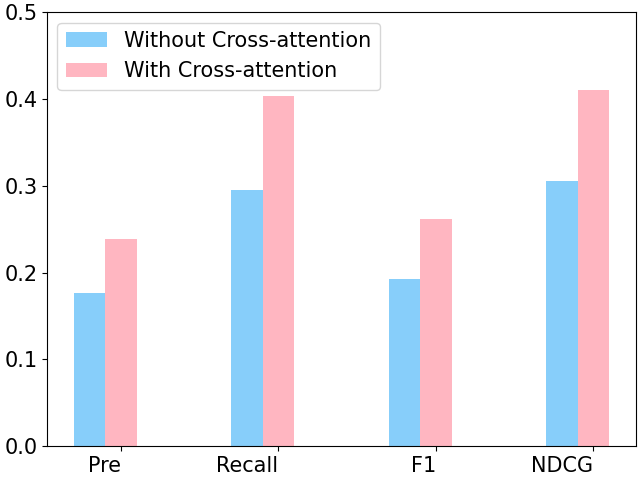}
    }
\caption{Performance and explanation evaluation w/o cross-attention.}
\label{fig: illustration of ablation study} 
\end{figure}

\subsection{Ablation Study}
In this subsection, we investigate the effectiveness of the cross-attention mechanism employed in our proposed method. Figure~\ref{fig: illustration of ablation study} presents the results of Amazon-Beauty dataset w/o employing the cross-attention mechanism. It can be seen that the cross-attention mechanism is essential to enhance the performance of both the recommendation evaluation and the explanation evaluation. Similar trends are observed in other datasets. For brevity, we only present the key results. 

\begin{table}[h]
\caption{Time cost for various methods across datasets}
\vspace{-1mm}
\centering
\scalebox{0.8} 
{
\label{tab: computational cost}
\begin{tabular}{cccccc}
\hline
\textbf{Methods}& \textbf{Instruments}& \textbf{Video}& \textbf{Beauty}&\textbf{Clothing}& \textbf{Explanations}\\
\hline 
NCF & 161' & 683' & 1006' & 1017' &  \xmark \\
VBPR & 153' & 495' & 987' & 1042' &  \xmark \\
CER & 309' & 1052' & 1471' & 1966' & \cmark \\
NAR & 569' & 942' & 1331' & 2336' & \cmark \\
MANAS & 4373' & 8297' & 132405' & 104453' & \xmark \\
\hline
Ours & 2532' & 4183'  & 9539' & 10343' & \cmark \\
\hline
\end{tabular}
}
\vspace{-3mm}
\end{table}
\subsection{Computational Cost Analysis}
In this subsection, we analyze the computational cost of our proposed method. The primary expense of our method lies in the bi-level architecture search. Specifically, the intra-layer search space comprises 4 choices, and the inter-layer search space consists of 18 choices, resulting in a total of 4 × 18 = 72 potential architectures. However, rather than performing an exhaustive discrete search, we relax the discrete space into a continuous, differentiable one. This relaxation allows us to employ gradient-based optimization methods, which are markedly more efficient than traditional techniques such as reinforcement learning or evolutionary algorithms.

Table~\ref{tab: computational cost} summarizes the running times of each method. While the bi-level search introduces some computational overhead during the search phase, the continuous relaxation substantially reduces this burden. It leads to faster overall running times compared to conventional NAS techniques, such as MANAS.


\begin{table}[h]
\caption{Obtained cross-attention and interaction functions}
\centering
\scalebox{0.9} 
{
\label{tab: obtained architecture}
\begin{tabular}{cccc}
\hline
\textbf{Datasets}& \textbf{Inputs}& \textbf{Intra-layer}& \textbf{Inter-layer}\\
\hline 
Instrument &$\tilde{f_u}, \tilde{f_i}, e_i$  & $\mathrm{att}_1$  &$\mathrm{Plus}$ \\
Video &$\tilde{f_u}, \tilde{f_i}, e_i$  &$\mathrm{att}_2$  &$\mathrm{Concat}$  \\
Beauty &$\tilde{f_u}, \tilde{f_i}$  & $\mathrm{att}_2$  &$\mathrm{Concat}$ \\
Clothing &$\tilde{f_u}, \tilde{f_i}, e_i, e_u$  &$\mathrm{att}_2$  & $\mathrm{Concat}, \mathrm{Multiply}$\\
\hline
\end{tabular}
}
\vspace{-3mm}
\end{table}

\subsection{Cross Attention and Interaction Functions Obtained}
In this subsection, we show the cross-attention and interaction functions that are obtained by our proposed method in Table ~\ref{tab: obtained architecture}.
We observed that utilizing three inputs yields better performance on the small-scale datasets. Additionally, the construction of the cross-attention mechanism exhibits slight variations across different datasets. 
Furthermore, 
there is no definitive winner among all operations, and the optimal operation may depend on the dataset. 
These observations demonstrate that no single architecture consistently outperforms the others, and there is no optimal architecture that works for all scenarios, supporting the notion that a universally optimal architecture is not feasible.
These observations further underscore the necessity for our proposed method, which dynamically adjusts the architecture to match the specific needs and challenges of various datasets. This adaptability ensures that the recommendation model remains robust and effective across diverse contexts.



\section{Related Work}
In this section, we briefly introduce relevant studies, focusing on explainable recommender systems and neural architecture search for recommendations.

\textbf{Explainable Recommender Systems.} 
Recent work has highlighted the importance of building reliable and trustworthy systems~\cite{wu2023towards, wu2024towards}. Explanations for recommendations are essential to enhance transparency, trust, and user engagement.
While traditional models like collaborative filtering~\cite {schafer2007collaborative} and matrix factorization~\cite{koren2009matrix} offer generic explanations, they often fail to capture individual user preferences. To address this, more advanced methods have emerged, including sentiment-aware models that analyze user reviews~\cite{zhang2014explicit, he2015trirank}, deep learning models like DeepCoNN that encode reviews for explainability~\cite{zheng2017joint, catherine2017transnets}, and aspect-based methods that select informative sentences using attention mechanisms.
Some approaches employ path-based techniques~\cite{geng2022path, zheng2022explainable}, while others utilize attention mechanisms~\cite{dong2020asymmetrical, barkan2020explainable}. 
Despite these advancements, designing attention architectures and path-based explanations requires extensive human effort.
LLMs have recently demonstrated remarkable versatility in various tasks~\cite{zeng2025vision, wang2025genuine}. However, directly relying on LLMs for explanation generation can lead to issues such as hallucination, factual inaccuracies, and inconsistencies~\cite{lubos2024llm}.
To address these challenges, we leverage NAS to automate architecture search, optimizing cross-attention structures and interaction functions. We then integrate LLMs to enhance explanation generation by the aligned features shared between users and items.

\textbf{Neural Architecture Search.} 
Designing effective neural architectures remains highly dependent on expert knowledge, making it challenging for non-experts and limiting the exploration of novel architectures. To mitigate this, NAS automates model design by optimizing network structures while minimizing human intervention~\cite{ren2021comprehensive}. The success of Zoph’s work~\cite{DBLP:conf/iclr/ZophL17} validated the feasibility of automated architecture search, leading to research focused on improving efficiency through parameter sharing~\cite{pham2018efficient}, modular search spaces~\cite{li2020block}, and differentiable search strategies~\cite{liu2018darts}.
With advancements in NAS, researchers have applied these techniques to recommendation systems. AutoField~\cite{wang2022autofield} automatically selects essential feature fields, while AutoCross~\cite{luo2019autocross} discovers high-order feature interactions in a structured search space. AutoFIS~\cite{liu2020autofis} identifies important feature interactions through an exhaustive enumeration approach, and AutoLoss~\cite{zhao2021autoloss} adapts loss functions for optimal model performance. More recently, NASRec~\cite{zhang2023nasrec} introduced weight-sharing supernets to generate diverse recommendation architectures, while Meta's latest studies emphasize the potential of NAS in ranking and recommendation tasks~\cite{wen2023rankitect, yin2023automl}.
Despite the growing adoption of NAS in recommendations, little research has focused on improving model explainability. Most existing NAS approaches prioritize predictive accuracy, often overlooking the need for transparent and interpretable recommendations. To bridge this gap, we propose Bi-NAS, a novel approach that searches for optimal cross-attention mechanisms and interaction functions, enhancing both recommendation accuracy and explainability.

\section{Conclusion}
The significance of explanations in recommender systems cannot be overstated. They serve as a critical link between users and recommended items, enhancing decision-making, building trust, and fostering user loyalty. Effective explanations can clarify the rationale behind algorithmic decisions and make the system more transparent and user-friendly.
However, traditional explainable models aimed at providing explanations usually fail to achieve the desired level of interpretability that the users seek, and few studies are dedicated to quantifying the effectiveness of interpretable recommendation systems.
Additionally, most current methods require substantial time and expert knowledge for the intricate construction of explainable pathways, attention mechanisms, etc. While LLMs demonstrate notable versatility in generating natural language text, directly using them for explanation generation can introduce issues such as hallucination, factual inaccuracies, a lack of true personalization, and inconsistencies.
To address these challenges, this paper introduces a Bi-NAS framework. By jointly optimizing cross-attention structures and feature interaction functions, Bi-NAS promotes the generation of effective explanations and minimizes the dependency on expert knowledge. Furthermore, LLMs have been employed to enhance explanation generation from the aligned features shared between users and items and user history, enabling more personalized justifications.
Our extensive experimental results demonstrate the efficacy of our method in enhancing recommender systems. This approach yields superior recommendations for the users while generating more effective explanations. 

\section*{Acknowledgements}
We thank the anonymous reviewers for their constructive comments. This work is supported by the National Science Foundation under Award No. IIS-2339989 and No. 2406439, DARPA under contract No. HR00112490370 and No. HR001124S0013, U.S. Department of Homeland Security under Grant Award No. 17STCIN00001-08-00,  Amazon-Virginia Tech Initiative for Efficient and Robust Machine Learning, Amazon AWS, Google, Cisco, 4-VA, Commonwealth Cyber Initiative, National Surface Transportation Safety Center for Excellence, and Virginia Tech. The views and conclusions are those of the authors and should not be interpreted as representing the official policies of the funding agencies or the government.

\bibliographystyle{IEEEtranN}
\footnotesize
\bibliography{sample-base}

@incollection{schafer2007collaborative,
  title={Collaborative filtering recommender systems},
  author={Schafer, J Ben and Frankowski, Dan and Herlocker, Jon and Sen, Shilad},
  booktitle={The adaptive web},
  pages={291--324},
  year={2007},
  publisher={Springer}
}

@article{koren2009matrix,
  title={Matrix factorization techniques for recommender systems},
  author={Koren, Yehuda and Bell, Robert and Volinsky, Chris},
  journal={Computer},
  volume={42},
  number={8},
  pages={30--37},
  year={2009},
  publisher={IEEE}
}

@inproceedings{balog2020measuring,
  title={Measuring recommendation explanation quality: The conflicting goals of explanations},
  author={Balog, Krisztian and Radlinski, Filip},
  booktitle={Proceedings of the 43rd international ACM SIGIR conference on research and development in information retrieval},
  pages={329--338},
  year={2020}
}

@inproceedings{xian2021ex3,
  title={Ex3: Explainable attribute-aware item-set recommendations},
  author={Xian, Yikun and Zhao, Tong and Li, Jin and Chan, Jim and Kan, Andrey and Ma, Jun and Dong, Xin Luna and Faloutsos, Christos and Karypis, George and Muthukrishnan, Shan and others},
  booktitle={Proceedings of the 15th ACM Conference on Recommender Systems},
  pages={484--494},
  year={2021}
}

@inproceedings{zheng2022explainable,
  title={Explainable session-based recommendation with meta-path guided instances and self-attention mechanism},
  author={Zheng, Jiayin and Mai, Juanyun and Wen, Yanlong},
  booktitle={Proceedings of the 45th International ACM SIGIR Conference on Research and Development in Information Retrieval},
  year={2022}
}

@inproceedings{barkan2020explainable,
  title={Explainable recommendations via attentive multi-persona collaborative filtering},
  author={Barkan, Oren and Fuchs, Yonatan and Caciularu, Avi and Koenigstein, Noam},
  booktitle={Proceedings of the 14th ACM Conference on Recommender Systems},
  pages={468--473},
  year={2020}
}

@inproceedings{geng2022path,
  title={Path language modeling over knowledge graphsfor explainable recommendation},
  author={Geng, Shijie and Fu, Zuohui and Tan, Juntao and Ge, Yingqiang and De Melo, Gerard and Zhang, Yongfeng},
  booktitle={Proceedings of the ACM Web Conference 2022},
  pages={946--955},
  year={2022}
}

@inproceedings{dong2020asymmetrical,
  title={Asymmetrical hierarchical networks with attentive interactions for interpretable review-based recommendation},
  author={Dong, Xin and Ni, Jingchao and Cheng, Wei and Chen, Zhengzhang and Zong, Bo and Song, Dongjin and Liu, Yanchi and Chen, Haifeng and De Melo, Gerard},
  booktitle={Proceedings of the AAAI conference on artificial intelligence},
  volume={34},
  number={05},
  pages={7667--7674},
  year={2020}
}

@inproceedings{zheng2017joint,
  title={Joint deep modeling of users and items using reviews for recommendation},
  author={Zheng, Lei and Noroozi, Vahid and Yu, Philip S},
  booktitle={Proceedings of the tenth ACM international conference on web search and data mining},
  pages={425--434},
  year={2017}
}

@inproceedings{wang2020attention,
  title={Attention-guide walk model in heterogeneous information network for multi-style recommendation explanation},
  author={Wang, Xin and Wang, Ying and Ling, Yunzhi},
  booktitle={Proceedings of the AAAI Conference on Artificial Intelligence},
  volume={34},
  number={04},
  pages={6275--6282},
  year={2020}
}

@inproceedings{chen2021temporal,
  title={Temporal meta-path guided explainable recommendation},
  author={Chen, Hongxu and Li, Yicong and Sun, Xiangguo and Xu, Guandong and Yin, Hongzhi},
  booktitle={Proceedings of the 14th ACM international conference on web search and data mining},
  pages={1056--1064},
  year={2021}
}

@article{ren2021comprehensive,
  title={A comprehensive survey of neural architecture search: Challenges and solutions},
  author={Ren, Pengzhen and Xiao, Yun and Chang, Xiaojun and Huang, Po-Yao and Li, Zhihui and Chen, Xiaojiang and Wang, Xin},
  journal={ACM Computing Surveys (CSUR)},
  volume={54},
  number={4},
  year={2021},
  publisher={ACM New York, NY, USA}
}

@inproceedings{DBLP:conf/iclr/ZophL17,
  author       = {Barret Zoph and
                  Quoc V. Le},
  title        = {Neural Architecture Search with Reinforcement Learning},
  booktitle    = {5th International Conference on Learning Representations, {ICLR} 2017, Toulon, France, April 24-26, 2017, Conference Track Proceedings},
  year         = {2017},
  bibsource    = {dblp computer science bibliography, https://dblp.org}
}

@inproceedings{pham2018efficient,
  title={Efficient neural architecture search via parameters sharing},
  author={Pham, Hieu and Guan, Melody and Zoph, Barret and Le, Quoc and Dean, Jeff},
  booktitle={International conference on machine learning},
  pages={4095--4104},
  year={2018},
  organization={PMLR}
}

@inproceedings{zoph2018learning,
  title={Learning transferable architectures for scalable image recognition},
  author={Zoph, Barret and Vasudevan, Vijay and Shlens, Jonathon and Le, Quoc V},
  booktitle={Proceedings of the IEEE conference on computer vision and pattern recognition},
  year={2018}
}

@inproceedings{li2020block,
  title={Block-wisely supervised neural architecture search with knowledge distillation},
  author={Li, Changlin and Peng, Jiefeng and Yuan, Liuchun and Wang, Guangrun and Liang, Xiaodan and Lin, Liang and Chang, Xiaojun},
  booktitle={Proceedings of the IEEE/CVF Conference on Computer Vision and Pattern Recognition},
  pages={1989--1998},
  year={2020}
}

@article{liu2018darts,
  title={Darts: Differentiable architecture search},
  author={Liu, Hanxiao and Simonyan, Karen and Yang, Yiming},
  journal={arXiv preprint arXiv:1806.09055},
  year={2018}
}

@inproceedings{ni2019justifying,
  title={Justifying recommendations using distantly-labeled reviews and fine-grained aspects},
  author={Ni, Jianmo and Li, Jiacheng and McAuley, Julian},
  booktitle={Proceedings of the 2019 Conference on Empirical Methods in Natural Language Processing and the 9th International Joint Conference on Natural Language Processing (EMNLP-IJCNLP)},
  pages={188--197},
  year={2019}
}

@inproceedings{he2017neural,
  title={Neural collaborative filtering},
  author={He, Xiangnan and Liao, Lizi and Zhang, Hanwang and Nie, Liqiang and Hu, Xia and Chua, Tat-Seng},
  booktitle={Proceedings of the 26th international conference on world wide web},
  pages={173--182},
  year={2017}
}

@inproceedings{he2016vbpr,
  title={VBPR: visual bayesian personalized ranking from implicit feedback},
  author={He, Ruining and McAuley, Julian},
  booktitle={Proceedings of the AAAI conference on artificial intelligence},
  volume={30},
  number={1},
  year={2016}
}

@article{zhou2021intrinsic,
  title={From intrinsic to counterfactual: On the explainability of contextualized recommender systems},
  author={Zhou, Yao and Wang, Haonan and He, Jingrui and Wang, Haixun},
  journal={arXiv preprint arXiv:2110.14844},
  year={2021}
}

@article{zhou2024based,
  title={Based Explainable Recommendations: A Transparency Perspective},
  author={Zhou, Yao and Wang, Haonan and He, Jingrui and Wang, Haixun},
  journal={ACM Transactions on Recommender Systems},
  year={2024},
  publisher={ACM New York, NY}
}

@inproceedings{chen2022learn,
  title={Learn basic skills and reuse: Modularized adaptive neural architecture search (manas)},
  author={Chen, Hanxiong and Li, Yunqi and Zhu, He and Zhang, Yongfeng},
  booktitle={Proceedings of the 31st ACM International Conference on Information \& Knowledge Management},
  pages={169--179},
  year={2022}
}

@inproceedings{zhang2014explicit,
  title={Explicit factor models for explainable recommendation based on phrase-level sentiment analysis},
  author={Zhang, Yongfeng and Lai, Guokun and Zhang, Min and Zhang, Yi and Liu, Yiqun and Ma, Shaoping},
  booktitle={Proceedings of the 37th international ACM SIGIR conference on Research \& development in information retrieval},
  pages={83--92},
  year={2014}
}

@inproceedings{zhang2014users,
  title={Do users rate or review? Boost phrase-level sentiment labeling with review-level sentiment classification},
  author={Zhang, Yongfeng and Zhang, Haochen and Zhang, Min and Liu, Yiqun and Ma, Shaoping},
  booktitle={Proceedings of the 37th international ACM SIGIR conference on Research \& development in information retrieval},
  pages={1027--1030},
  year={2014}
}

@inproceedings{tan2021counterfactual,
  title={Counterfactual explainable recommendation},
  author={Tan, Juntao and Xu, Shuyuan and Ge, Yingqiang and Li, Yunqi and Chen, Xu and Zhang, Yongfeng},
  booktitle={Proceedings of the 30th ACM International Conference on Information \& Knowledge Management},
  pages={1784--1793},
  year={2021}
}

@inproceedings{zhao2021autoloss,
  title={Autoloss: Automated loss function search in recommendations},
  author={Zhao, Xiangyu and Liu, Haochen and Fan, Wenqi and Liu, Hui and Tang, Jiliang and Wang, Chong},
  booktitle={Proceedings of the 27th ACM SIGKDD Conference on Knowledge Discovery \& Data Mining},
  pages={3959--3967},
  year={2021}
}

@inproceedings{yao2020efficient,
  title={Efficient neural interaction function search for collaborative filtering},
  author={Yao, Quanming and Chen, Xiangning and Kwok, James T and Li, Yong and Hsieh, Cho-Jui},
  booktitle={Proceedings of The web conference 2020},
  pages={1660--1670},
  year={2020}
}

@inproceedings{lin2022adafs,
  title={AdaFS: Adaptive feature selection in deep recommender system},
  author={Lin, Weilin and Zhao, Xiangyu and Wang, Yejing and Xu, Tong and Wu, Xian},
  booktitle={Proceedings of the 28th ACM SIGKDD Conference on Knowledge Discovery and Data Mining},
  pages={3309--3317},
  year={2022}
}

@article{dubey2024llama,
  title={The llama 3 herd of models},
  author={Dubey, Abhimanyu and Jauhri, Abhinav and Pandey, Abhinav and Kadian, Abhishek and Al-Dahle, Ahmad and Letman, Aiesha and Mathur, Akhil and Schelten, Alan and Yang, Amy and Fan, Angela and others},
  journal={arXiv preprint arXiv:2407.21783},
  year={2024}
}

@inproceedings{he2015trirank,
  title={Trirank: Review-aware explainable recommendation by modeling aspects},
  author={He, Xiangnan and Chen, Tao and Kan, Min-Yen and Chen, Xiao},
  booktitle={Proceedings of the 24th ACM international on conference on information and knowledge management},
  pages={1661--1670},
  year={2015}
}

@inproceedings{catherine2017transnets,
  title={Transnets: Learning to transform for recommendation},
  author={Catherine, Rose and Cohen, William},
  booktitle={Proceedings of the eleventh ACM conference on recommender systems},
  pages={288--296},
  year={2017}
}

@inproceedings{wang2022autofield,
  title={Autofield: Automating feature selection in deep recommender systems},
  author={Wang, Yejing and Zhao, Xiangyu and Xu, Tong and Wu, Xian},
  booktitle={Proceedings of the ACM Web Conference 2022},
  pages={1977--1986},
  year={2022}
}

@inproceedings{luo2019autocross,
  title={Autocross: Automatic feature crossing for tabular data in real-world applications},
  author={Luo, Yuanfei and Wang, Mengshuo and Zhou, Hao and Yao, Quanming and Tu, Wei-Wei and Chen, Yuqiang and Dai, Wenyuan and Yang, Qiang},
  booktitle={Proceedings of the 25th ACM SIGKDD International Conference on Knowledge Discovery \& Data Mining},
  year={2019}
}

@inproceedings{liu2020autofis,
  title={Autofis: Automatic feature interaction selection in factorization models for click-through rate prediction},
  author={Liu, Bin and Zhu, Chenxu and Li, Guilin and Zhang, Weinan and Lai, Jincai and Tang, Ruiming and He, Xiuqiang and Li, Zhenguo and Yu, Yong},
  booktitle={proceedings of the 26th ACM SIGKDD international conference on knowledge discovery \& data mining},
  pages={2636--2645},
  year={2020}
}

@inproceedings{abdollahi2016explainable,
  title={Explainable matrix factorization for collaborative filtering},
  author={Abdollahi, Behnoush and Nasraoui, Olfa},
  booktitle={Proceedings of the 25th International Conference Companion on World Wide Web},
  pages={5--6},
  year={2016}
}

@inproceedings{chen2019co,
  title={Co-attentive multi-task learning for explainable recommendation.},
  author={Chen, Zhongxia and Wang, Xiting and Xie, Xing and Wu, Tong and Bu, Guoqing and Wang, Yining and Chen, Enhong},
  booktitle={IJCAI},
  pages={2137--2143},
  year={2019}
}

@inproceedings{wang2018explainable,
  title={Explainable recommendation via multi-task learning in opinionated text data},
  author={Wang, Nan and Wang, Hongning and Jia, Yiling and Yin, Yue},
  booktitle={The 41st international ACM SIGIR conference on research \& development in information retrieval},
  pages={165--174},
  year={2018}
}

@article{wistuba2019survey,
  title={A survey on neural architecture search},
  author={Wistuba, Martin and Rawat, Ambrish and Pedapati, Tejaswini},
  journal={arXiv preprint arXiv:1905.01392},
  year={2019}
}

@inproceedings{chen2020try,
  title={Try this instead: Personalized and interpretable substitute recommendation},
  author={Chen, Tong and Yin, Hongzhi and Ye, Guanhua and Huang, Zi and Wang, Yang and Wang, Meng},
  booktitle={Proceedings of the 43rd international ACM SIGIR conference on research and development in information retrieval},
  pages={891--900},
  year={2020}
}

@book{aggarwal2016recommender,
  title={Recommender systems},
  author={Aggarwal, Charu C and others},
  volume={1},
  year={2016},
  publisher={Springer}
}

@inproceedings{zhang2023nasrec,
  title={NASRec: weight sharing neural architecture search for recommender systems},
  author={Zhang, Tunhou and Cheng, Dehua and He, Yuchen and Chen, Zhengxing and Dai, Xiaoliang and Xiong, Liang and Yan, Feng and Li, Hai and Chen, Yiran and Wen, Wei},
  booktitle={Proceedings of the ACM Web Conference 2023},
  pages={1199--1207},
  year={2023}
}

@article{wen2023rankitect,
  title={Rankitect: Ranking Architecture Search Battling World-class Engineers at Meta Scale},
  author={Wen, Wei and Liu, Kuang-Hung and Fedorov, Igor and Zhang, Xin and Yin, Hang and Chu, Weiwei and Hassani, Kaveh and Sun, Mengying and Liu, Jiang and Wang, Xu and others},
  journal={arXiv preprint arXiv:2311.08430},
  year={2023}
}

@article{yin2023automl,
  title={AutoML for Large Capacity Modeling of Meta Ranking Systems},
  author={Yin, Hang and Liu, Kuang-Hung and Sun, Mengying and Chen, Yuxin and Zhang, Buyun and Liu, Jiang and Sehgal, Vivek and Panchal, Rudresh Rajnikant and Hotaj, Eugen and Liu, Xi and others},
  journal={arXiv preprint arXiv:2311.07870},
  year={2023}
}

@article{wang2025genuine,
  title={GENUINE: Graph Enhanced Multi-level Uncertainty Estimation for Large Language Models},
  author={Wang, Tuo and Kulkarni, Adithya and Cody, Tyler and Beling, Peter A and Yan, Yujun and Zhou, Dawei},
  journal={arXiv preprint arXiv:2509.07925},
  year={2025}
}

@inproceedings{zeng2025vision,
  title={Are vision llms road-ready? a comprehensive benchmark for safety-critical driving video understanding},
  author={Zeng, Tong and Wu, Longfeng and Shi, Liang and Zhou, Dawei and Guo, Feng},
  booktitle={Proceedings of the 31st ACM SIGKDD Conference on Knowledge Discovery and Data Mining V. 2},
  pages={5972--5983},
  year={2025}
}

@inproceedings{wu2022towards,
  title={Towards high-order complementary recommendation via logical reasoning network},
  author={Wu, Longfeng and Zhou, Yao and Zhou, Dawei},
  booktitle={2022 IEEE International Conference on Data Mining (ICDM)},
  year={2022},
  organization={IEEE}
}

@inproceedings{lubos2024llm,
  title={LLM-generated explanations for recommender systems},
  author={Lubos, Sebastian and Tran, Thi Ngoc Trang and Felfernig, Alexander and Polat Erdeniz, Seda and Le, Viet-Man},
  booktitle={Adjunct Proceedings of the 32nd ACM Conference on User Modeling, Adaptation and Personalization},
  pages={276--285},
  year={2024}
}

@inproceedings{wu2023towards,
  title={Towards reliable rare category analysis on graphs via individual calibration},
  author={Wu, Longfeng and Lei, Bowen and Xu, Dongkuan and Zhou, Dawei},
  booktitle={Proceedings of the 29th ACM SIGKDD Conference on Knowledge Discovery and Data Mining},
  pages={2629--2638},
  year={2023}
}

@inproceedings{wu2024towards,
  title={Towards Trustworthy Graph Neural Networks and Their Applications in Recommender Systems},
  author={Wu, Longfeng},
  booktitle={2024 IEEE International Conference on Big Data (BigData)},
  pages={8250--8252},
  year={2024},
  organization={IEEE}
}

\end{document}